\documentclass[preprint,onecolumn]{revtex4}
\usepackage{graphicx}

\begin{document}

\title{Molecular Cloud Biology}
%\title{Nebula-Relay Hypothesis: Methanogens in Molecular Clouds?}
\author{Lei~Feng}
\email{fenglei@pmo.ac.cn}
\affiliation{Key Laboratory of Dark Matter and Space Astronomy, Purple Mountain Observatory, Chinese Academy of Sciences, Nanjing 210023}
\affiliation{School of Astronomy and Space Science, University of Science and Technology of China, Hefei, Anhui 230026, China}
\affiliation{Joint Center for Particle, Nuclear Physics and Cosmology,  Nanjing University -- Purple Mountain Observatory,  Nanjing  210093, China}

\begin{abstract}

Some astrobiological models suggest that molecular clouds may serve as habitats for extraterrestrial life. This study reviews recent theoretical work addressing the physical and biochemical prerequisites for life in such environments, with particular focus on three subjects: (1) bioenergetic pathways under extreme low-temperature conditions; (2) the emergence and preservation of biomolecular chirality; and (3) detection methodologies for potential biosignatures. In this paper, we formally introduce the molecular cloud biology concept, which integrates all physicochemical and metabolic processes hypothesized to sustain life within molecular clouds. As a potential branch of astrobiology, molecular cloud biology warrants interdisciplinary collaborative research to validate its foundational assumptions and explore its scientific implications.

\end{abstract}
\maketitle
%%%%%%%%%%%%%%%%%%%%%%%%%%%%%%%%%%%%%%%%%%%%%%%%%%%%%%%%%%%%%%%%%%%
\section{Motivations}

There are mainly two competing models regarding the origin of life on Earth: abiogenesis~\cite{Oparin,Haldane} and panspermia~\cite{Panspermia}. The panspermia hypothesis was initially proposed by distinguished scientists such as Lord Kelvin and Hermann von Helmholtz. This model suggests that primitive life wandered through interstellar space before being transported to Earth. Notably, some research has demonstrated that microbes have the potential capacity to withstand extreme impact forces and hypervelocity conditions associated with interplanetary transfer~\cite{Rampelotto}.

Molecular clouds, primarily composed of gas and dust, are distributed within the interstellar medium of galaxies. They originate from the remnants of supernova explosions of earlier generations of stars and serve as the sites of star formation. Typically, the temperature of molecular clouds ranges from several Kelvin to a few dozen Kelvin, with an average density of approximately $\rm 10^2$ to $\rm 10^4$ molecules per cubic centimeter, where hydrogen molecules are the most abundant components. Additionally, molecular clouds contain a variety of organic compounds, including fullerenes \cite{fullerenes}, polycyclic aromatic hydrocarbons \cite{hydrocarbons}, glycolaldehyde \cite{glycolaldehyde}, and others. The panspermia hypothesis suggests that life forms could have traveled through and survived within molecular clouds. Consequently, investigating the biophysical and biochemical processes of life in these environments represents significant scientific interest.

Additionally, the nebula-relay hypothesis proposes that life's precursors originated in the planetary system of the Sun's stellar progenitor and subsequently experienced cosmic evolution through the pre-solar nebula \cite{nebula-relay}. The evolutionary timeline of terrestrial life can be divided into three discrete epochs: (1) primordial biopoiesis within the progenitor star's planetary environment, (2) a nebular incubation phase that sustained protobiological systems, and (3) subsequent biological evolution on Earth. The maintenance and development of Earth life's precursors under the unique physicochemical conditions of molecular clouds constitute a subject of particular astrobiological interest.

In the temperature range of $\rm 13.99-20.27 K$, hydrogen molecules remain in a liquid state under suitable pressure. Interestingly, these temperatures align with the typical conditions of molecular clouds, which presents a compelling coincidence. If life within molecular clouds were to possess hydrogen-enriched cell-like membrane structures, the internal environment of molecular clouds could potentially contain liquid hydrogen, analogous to the aqueous environment essential for terrestrial life. However, the low density of hydrogen molecules in molecular clouds makes their accumulation highly improbable. Notably, hydrogen molecules differ fundamentally from water molecules in their physicochemical properties. For instance, hydrogen molecules are nonpolar, a characteristic that is intrinsically less conducive to the biochemical complexity required for life. Could a portion of water molecules undergo phase transition to liquid hydrogen under the cryogenic temperatures and high-radiation conditions prevailing in such environments? Furthermore, is it thermodynamically feasible for a heterogeneous system-comprising solid water (ice) and liquid hydrogen-to coexist? Given these unresolved questions, we propose this provocative hypothesis while deferring a rigorous exploration of its technical complexities.

The energy sustaining terrestrial life predominantly originates from solar radiation. In contrast, within molecular cloud environments, what physicochemical mechanisms could supply the requisite energy to sustain biological processes? We propose that cosmic rays may serve as a viable energy sources, either through catalytic chemical pathways \cite{Methanogens} or by harnessing ionization energy generated by cosmic-ray interactions \cite{cosmic-ray-bioenergetics}. This hypothesis will be systematically analyzed in Sections II and III.

We have quantitatively analyzed the probability of homochiral polymer formation in cryogenic molecular clouds, demonstrating that that low-temperature environments may facilitate the generation of chiral polymer chains \cite{Chirality}. This subject is discussed in Section IV. If the primitive life in the pre-solar nebula had been dispersed among various celestial bodies in the solar system during the formation of the solar system, these primitive life forms and/or their fossilized remnants are predicted to be widely distributed across the solar system. Furthermore, prebiotic processes within molecular clouds could generate detectable biosignatures. The methods for investigating and verifying molecular cloud biology will be systematically discussed in Section V. A comprehensive summary and conclusions are presented in the final section.

\section{Possibilities for methanogenic and acetogenic life in molecular cloud}

On Earth, methanogenic archaea employ methanogenesis - a unique anaerobic respiratory mechanism - to generate Gibbs free energy through $\rm CO2$-to-$\rm CH4$  conversion. In \cite{McKay2005}, McKay and Smith explored the possibilities of methanogenic life on Titan. Since analogous molecules also exist in molecular clouds, we propose that methanogenesis processes could sustain life in these interstellar regions.

\subsection{The Gibbs free energy released from methanogenic processes}
The Gibbs free energy ($\Delta G$) is released from the following biochemical reaction for molecular cloud life
\begin{eqnarray}
\rm{CO_2+4H_2\rightarrow CH_4+2H_2O},\nonumber\\
\rm{2CO_2+4H_2\rightarrow CH_3COOH+2H_2O},\\
\rm{CO+3H_2\rightarrow CH_4+H_2O},\nonumber\\
\rm{C_2H_2+3H_2\rightarrow 2CH_4}.\nonumber
\end{eqnarray}
The common feature of these reactions is that the electron donors are all hydrogen molecules and the electron donors are various compounds of carbon.

%Following Ref.~\cite{McKay2005},
We calculate the Gibbs free energy with the method described in Ref. \cite{Kral1998} as follows
\begin{equation}
\Delta G=\Delta H-T\Delta S+RT{\rm ln}(Q),
\end{equation}
where $\Delta H$ ($\Delta S$) is the difference in heats of formation (entropy) between the products and reactants, $ R$ is the universal gas constant and $ T$ is the temperature of molecular cloud. $Q$ is the ratio of the activities which equal to their partial pressure for gas molecules and equal to unity for solid carbon dioxide.

\begin{figure}[htbp]
\centering
\includegraphics[width=0.7\textwidth]{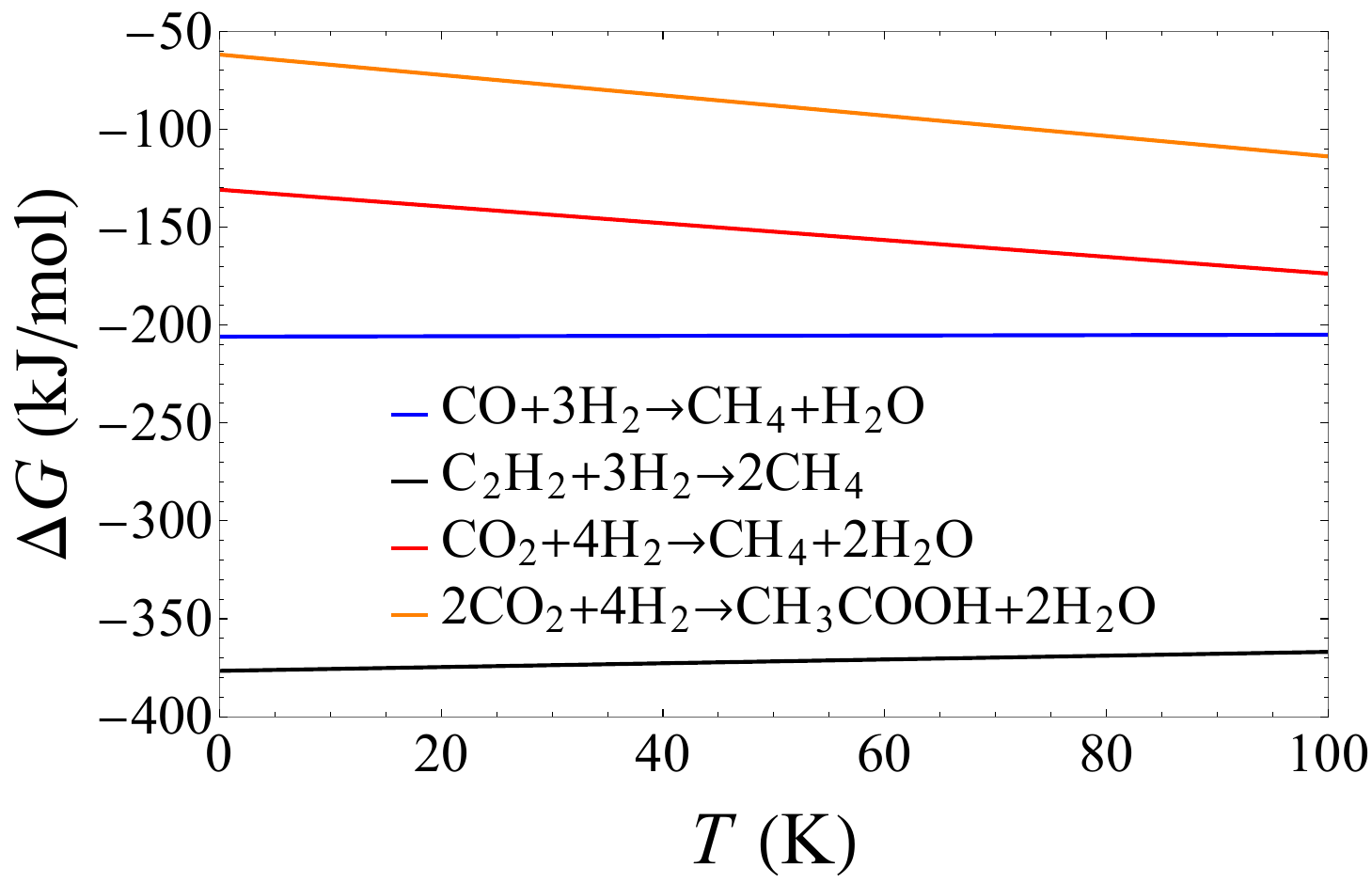}
\caption{The Gibbs free energy released from the synthesis of hydrocarbons \cite{Methanogens}.}
\label{fig:deltaG}
\end{figure}

The Gibbs free energy released by the above reactions is shown in Fig. \ref{fig:deltaG} and ranges from approximately $\rm -60$ to $\rm -370~kJ/mol$ \cite{Methanogens}. The negative sign indicates the release of energy. The minimum energy required for the survival of methanogens is approximately $\rm 42 kJ/mol$ on Earth \cite{Kral1998}, whereas this value may be lower in low-temperature environments. Therefore, the released Gibbs free energies are sufficient to sustain life activity in the molecular cloud.

%The density of acetylene in molecular clouds is remarkably low, significantly constraining the potential energy production rate. While carbon monoxide exhibits a higher density, its capacity to provide sufficient free energy depends critically on the intensity of potential life processes within the molecular cloud environment. Notably, carbon dioxide primarily exists in a solid state rather than as a gas phase in these regions~\cite{Minh1988,Gerakines,Whittet}. This solid  carbon dioxide represents a concentrated carbon source and potential reactant for biochemical processes. Consequently, hypothetical methanogenic or acetogenic life forms might potentially develop mechanisms to utilize these solid carbon dioxide deposits, possibly through surface attachment or other adaptive strategies.

The density of acetylene in molecular clouds is exceptionally low, which severely limits the potential energy production rate. Although carbon monoxide has a higher density, its capacity to provide sufficient free energy heavily depends on the activity level of hypothetical life processes within molecular cloud environments. Interestingly, carbon dioxide predominantly exists in the solid phase rather than the gaseous phase in these regions~\cite{Minh1988,Gerakines,Whittet}. This solid carbon dioxide serves as both a concentrated carbon source and a viable reactant for biochemical processes. Therefore, hypothetical methanogenic or acetogenic life forms could develop mechanisms to utilize these solid $\rm CO_2$ deposits, potentially through surface adsorption or other adaptive strategies.

\subsection{The Relationship with LUCA}

If methanogenic or acetogenic life forms existed in the pre-solar nebula and subsequently reached early Earth, they could potentially represent extraterrestrial precursors to terrestrial life. Intriguingly, the primitive atmosphere of early Earth was characterized by high concentrations of carbon monoxide and carbon dioxide \cite{KASTING,KASTING2,Holland,Miller1974}. This carbon-rich environment would have provided an optimal habitat for the survival and proliferation of methanogenic and acetogenic life forms. Some tentative evidence from previous studies \cite{Pierce,Rother} demonstrates that certain acetogenic and methanogenic archaea actively utilized carbon monoxide as a primary carbon source during Earth's early evolutionary stages.

More broadly, the Last Universal Common Ancestor (LUCA) is hypothesized to have been a life form that used $\rm H_2$, either $\rm CO$ or $\rm CO_2$, as electron donors or acceptors in its energy-generating chemical reactions. Some tentative evidence supports this hypothesis. Recent studies suggest that early life forms may have originated through methanogenesis in cooler environments, later adapting to hydrothermal vent habitats and giving rise to LUCA. This hypothesis finds support from the reconstruction of LUCA's minimal genome \cite{mat}. Furthermore, genomic analyses of nearly two thousand modern microbial species reveal that LUCA's genetic markers strongly indicate methanogenic and acetogenic origins \cite{Madeline}, providing tentative evidence for this evolutionary pathway.

\section{Cosmic ray-driven bioenergetics}

In bioenergetics, energetic electrons are gradually released through the electron transport chain. Moreover, these energetic electrons might be produced by cosmic ray ionization process, which suggests that cosmic ray-induced hydrogen ionization may potentially energize the electron transport chain. In Ref. \cite{Padovani:2009bj}, the authors found that the flux of cosmic ray protons and electrons is significantly enhanced at lower energy ranges ($\rm \leq 1 GeV$), which is favorable for producing biologically relevant electrons. This mechanism could potentially sustain the electron transport chain in biological systems.

%%%%%%%%%%%%%%%%%%%%%%%%%%%%%%%%%%%%%%%%%%%%%%%%%%%%%%%%%%%%%Figure%%%%%%%%%%%%%%%%%%%%%%%%%%%%%%%%%%%%%%%%%%%%%%%%%%%%%%%%%%%%%%%%%%%%%%%%
\begin{figure}[htbp]
\centering
\includegraphics[width=0.9\textwidth]{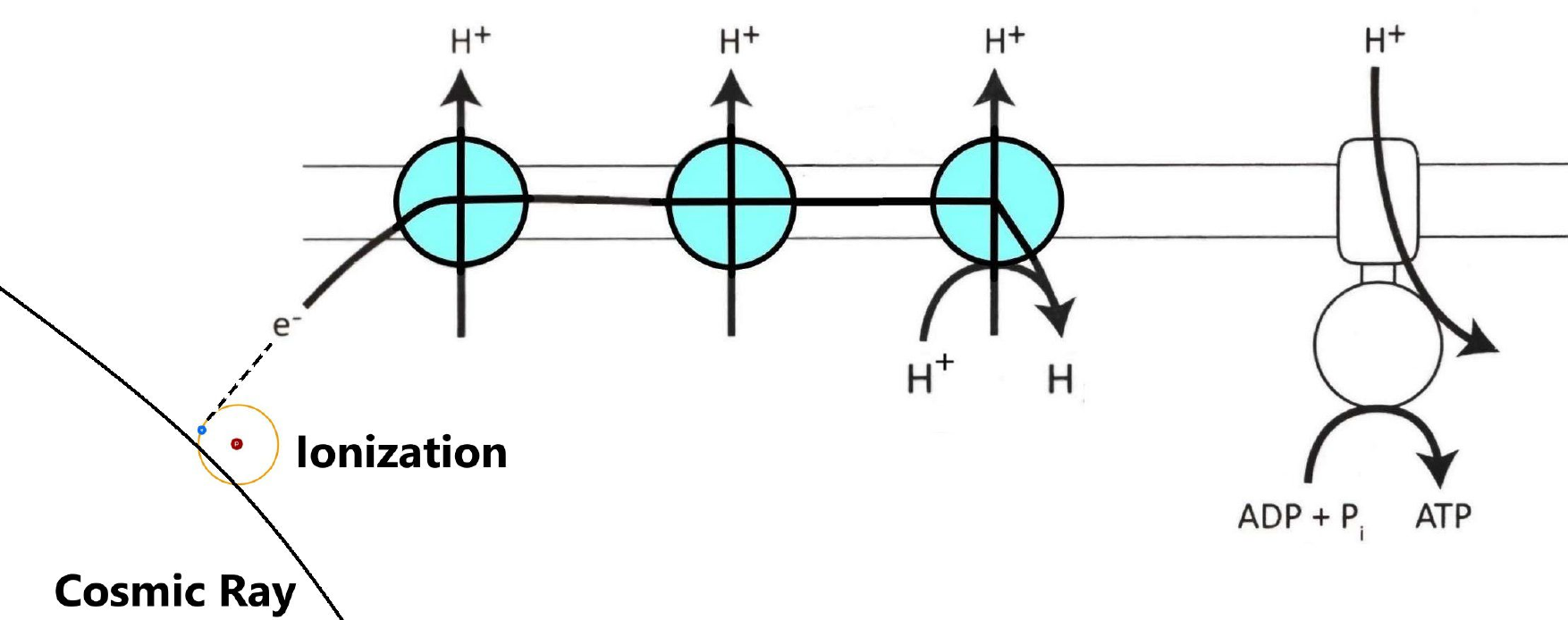}
\caption{Schematic diagram of the electron transport chain powered by the CR ionization of hydrogen where ATP and ADP are the abbreviation of adenosine triphosphate and adenosine di-phosphate respectively \cite{cosmic-ray-bioenergetics}.}
\label{fig:chain}
\end{figure}
%%%%%%%%%%%%%%%%%%%%%%%%%%%%%%%%%%%%%%%%%%%%%%%%%%%%%%%%%%%%%%%%%%%%%%%%%%%%%%%%%%%%%%%%%%%%%%%%%%%%%%%%%%%%%%%%%%%%%%%%%%%%%%%%%%%%%%%%%%%

The ionization of hydrogen can be discribed by the following processes
\begin{eqnarray}
{\rm CR} + {\rm H_2} &\rightarrow& {\rm CR} + {\rm H^+_2} + {\rm e}, \nonumber\\
{\rm CR} + {\rm H_2} &\rightarrow& {\rm CR} + {\rm H} + {\rm H^+} + {\rm e},\\
{\rm CR} + {\rm H_2} &\rightarrow& {\rm CR} + {\rm 2H^+} + {\rm 2e}, \nonumber
\end{eqnarray}
where ${\rm CR}$ denotes the charged cosmic ray particles.
The produced ion ${\rm H^+_2}$ will be depleted rapidly by reacting with hydrogen molecule through the following reaction \cite{ionreaction}
\begin{equation}
{\rm H^+_2} + {\rm H_2} \rightarrow {\rm H^+_3} + {\rm H}.
\end{equation}
Then ${\rm H^+_3}$ is removed by following complex reactions, and more details can be found in Ref. \cite{cosmic-ray-bioenergetics}. However, the produced secondary electrons may still have very high energy and can not be directly involved in the electron transport chain. The ionization process will continue and induce much more electrons with proper energy.

The proposed mechanism suggests that cosmic ray-driven energy transduction operates analogously to biological chemiosmosis, where a transmembrane proton motive force facilitates the phosphorylation of nucleotide triphosphates via rotary ATP synthase-like complexes. As depicted in Fig. \ref{fig:chain} \cite{cosmic-ray-bioenergetics}, the electron transport chain culminates in the regeneration of hydrogen atoms, thereby establishing a self-sustaining redox cycle. Through this process, cosmic ray energy is gradually released and ultimately converted into chemical energy that can be utilized by life forms. This non-canonical bioenergetic pathway uniquely utilizes hydrogen as both electron donors ($\rm H_2$) and terminal acceptors ($\rm H^+$), in contrast to Earth's oxygen-dominated electron transport chains.

Given the scarcity of organic molecules and the abundance of hydrogen molecules in molecular clouds, developing a proton gradient is a natural way to drive energy conversion. This process may be fundamentally  related to the origin of chemiosmosis.

\section{The Chirality Puzzle of Biological Molecules}

Chirality is typically broken in the biochemical processes. For instance, proteins in nearly all terrestrial organisms exclusively consist of L-amino acids. The origin of Chirality is still an unresolved scientific puzzle. In Ref. \cite{Chirality}, the author calculated the amplification of left-right asymmetry by producing homochiral polymers following the method presented in Ref. \cite{luo1994}.

We first assume that the distribution of chiral monomers satisfies the Boltzmann distribution, and the polarization of chiral monomers can be described as follows
\begin{equation}
\eta=\left(n_{\rm L}-n_{\rm D}\right) /\left(n_{\rm L}+n_{\rm D}\right)={\rm tanh}~[\left(E_{\rm D}-E_{\rm L}\right) / 2 K_{\rm B} T]={\rm tanh}~J,
\end{equation}
where $K_{\rm B}$ is the Boltzmann constant and $E_{\rm L,D}$ is the energy of L(D)-form chiral monomer. Following the calculations in Ref. \cite{luo1994},
the chiral polarization of polymer $f_{L}$ can be calculated through a statistical ensemble with the Ising model \cite{ising}, which is
\begin{equation}
f_{\rm L} =\frac{1}{2}\left\{1+{\rm sinh} J /\left[{\rm sinh}^{2} J+e^{-4 U}\right]^{1 / 2}\right\},
\end{equation}
where $f_{\rm L(D)}$ denotes the probability of L(D)-form monomer in the chain and $U=\left(E_{\rm L D}-E_{\rm L L}\right) / 2 K_{\rm B} T$ which is the deviation of recombined energy between different types of monomers.

\begin{figure}[htbp]
\centering
\includegraphics[width=0.49\textwidth]{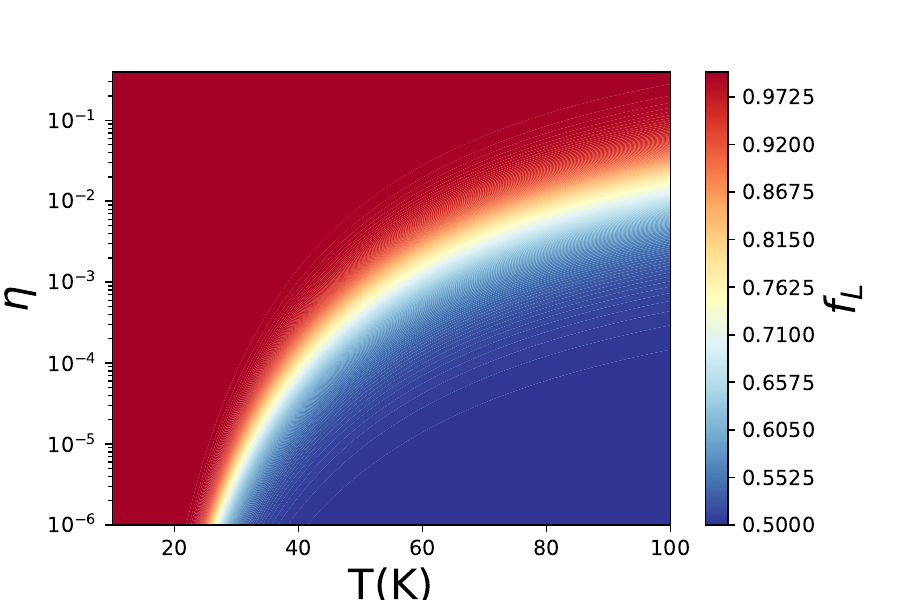}
\includegraphics[width=0.49\textwidth]{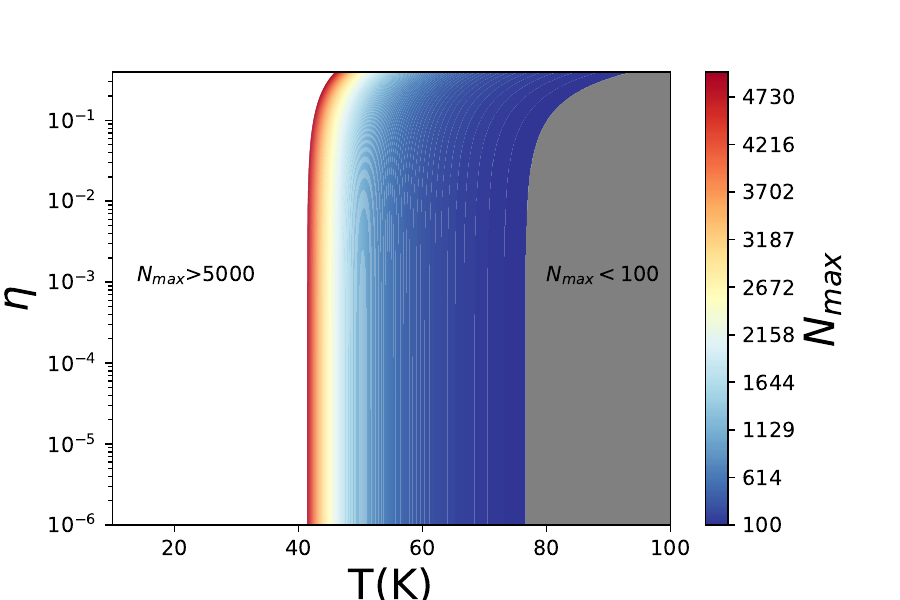}
\caption{The probability of L-type monomer in the polymer chain (left figure) and the dependence of the maximum pure L-type chain length N on $\eta$ and temperature (right figure).}
\label{fig:fl}
\end{figure}

The formation of the peptide chain is an exothermic reaction. If reactions involving homochiral molecules release more energy, purely chiral chains can be more readily induced. The probability of L-form monomer in the polymer chain is presented in Fig. \ref{fig:fl} \cite{Chirality}. The figure demonstrates that large polarization of chiral monomers and low temperature enlarge the probability of an L-form monomer. At low temperatures ($T<{\rm 25~K}$), even a minimal deviation between $\rm L-D$ type monomer could lead to the formation of left-handed chains. For $\eta \sim {\rm 10^{-2}}$, temperatures below $\rm 100 K$ are all acceptable.

We define the maximum permissible length of pure L-type polymer chain $N_{max}$ with the detailed calculation procedure provided in Ref.~\cite{luo1994}. The relationship between $N_{max}$, $\rm T$ and $\eta$ is shown in Fig. \ref{fig:fl}. From the figure, it is easy to see that $N_{max}$ is less than 100 for $T~>~{\rm 80~K}$ and larger than 5000 for $T~<~{\rm 40~K}$. These resultssuggest that pure left-handed chiral chains can be naturally synthesized under molecular cloud conditions.

\section{Testing Molecular Cloud Biology and Searching for Molecular Cloud Life}

In this section, we systematically discuss the methodological framework for investigating molecular cloud biology and the methods to detect and analyze potential descendants and fossilized remnants of molecular cloud life.

\subsection{Laboratory-based Simulation}

The direct detection of potential life forms within molecular clouds is almost impossible under current observational capabilities. However, we can systematically explore the fundamental biochemical processes of molecular cloud life forms through laboratory simulations that precisely replicate the extreme conditions characteristic of molecular clouds. These experimental approaches not only enable the examination of the viability and metabolic pathways of putative primitive life forms but also offer critical insights into their potential evolutionary adaptations.  This experimental methodology represents one of the most promising and direct approaches for enhancing our understanding of molecular cloud biology.

Although dedicated experiments specifically targeting molecular cloud conditions have yet to be conducted, several relevant studies with significant implications have been performed. Notably, in Ref.~\cite{Seager2020}, researchers demonstrated that both \textit{Escherichia coli} and yeast exhibit remarkable adaptability, maintaining robust growth and reproduction in pure hydrogen atmospheres. This insightful experiment, while primarily designed to simulate the atmospheric conditions of gaseous exoplanets, provides valuable and reference-worthy information on biological processes of molecular cloud life forms, given their similar hydrogen-dominated composition.

The bioenergetic mechanisms presented in Sections II and III can be experimentally validated through simulation experiments, as demonstrated in Refs. \cite{Methanogens,cosmic-ray-bioenergetics}. The formation of chiral polymer chains in low-temperature environments can also be simulated experimentally, as shown in Ref. \cite{Chirality}. Overall, the fundamental principles of molecular cloud biology can be systematically investigated and empirically validated using current laboratory technologies.

\subsection{Homologous extraterrestrial life within the solar system}

Based on the hypothesis that primitive molecular cloud life forms were captured and subsequently integrated into various celestial bodies during the formation of the solar system, these genetically related life forms may have already settled in habitable celestial bodies throughout the solar system. %This theoretical framework suggests that life forms in the pre-solar nebula may be the ancestors of terrestrial life.

Besides Earth, several celestial bodies within our solar system are suitable habitats for the survival and propagation of life forms. Europa and Enceladus are ideal candidates for testing our hypothesis. The thick ice sheets enveloping these celestial bodies have effectively shielded them from the continuous influx of external extraterrestrial seeds. Therefore, if life forms with DNA/RNA/protein systems are found on Europa or Enceladus, it would strongly suggest the existence of homologous life seeds potentially originating from molecular clouds.

Titan, a methane-rich moon, has been proposed as a potential host for methane-based life forms \cite{McKay2005}. The methanogenic life forms, hypothesized to originate from molecular clouds \cite{Methanogens}, could potentially thrive in Titan's environment. Additionally, comets may serve as effective carriers of  molecular cloud life, making them promising targets in the search for homologous extraterrestrial organisms.

\subsection{Biomarker molecules in molecular clouds}

The presence of life forms within molecular clouds would inevitably lead to the production of biologically significant molecules through metabolic activities. Consequently, the search for biomarker molecules in molecular clouds and protostellar environments emerges as a highly promising investigative approach. Given that protostars originate from molecular clouds, any existing life forms within these clouds would naturally become concentrated in these regions. This concentration effect would result in a corresponding increase in biomarker density, potentially making these areas optimal for detecting biosignatures and other indicators of extraterrestrial life.

The identification of potential biomarker molecules involves many kinds of chemical compounds. Conventional biosignatures, such as methane and phosphine, have long been recognized as reliable indicators of biological activity. Notably, as demonstrated in Ref. \cite{Pilcher}, organosulfur compounds have been confirmed as significant biomarkers for detecting extraterrestrial life. %These molecular species collectively form a robust set of indicators for astrobiological investigations within molecular clouds.

\begin{figure}[htbp]
\centering
\includegraphics[width=0.45\textwidth]{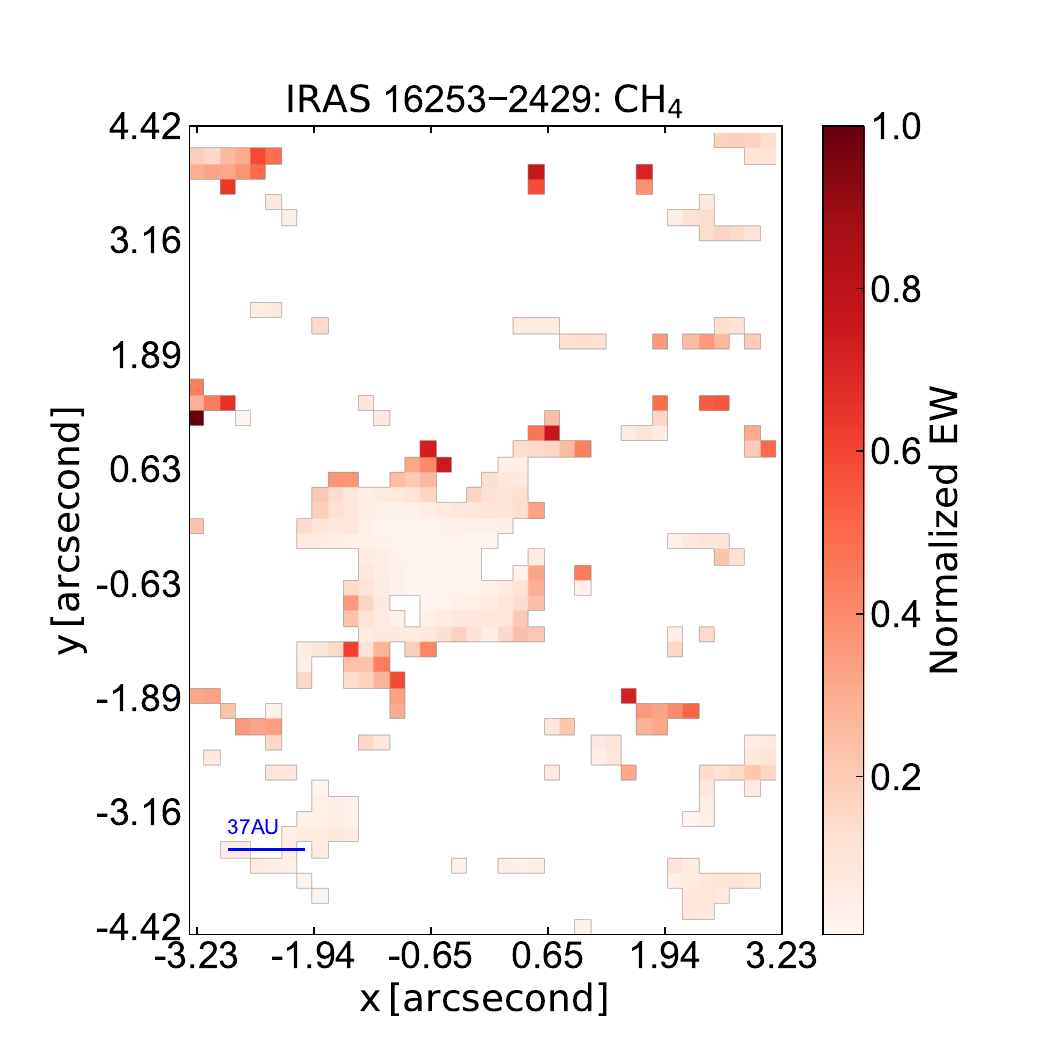}
\includegraphics[width=0.45\textwidth]{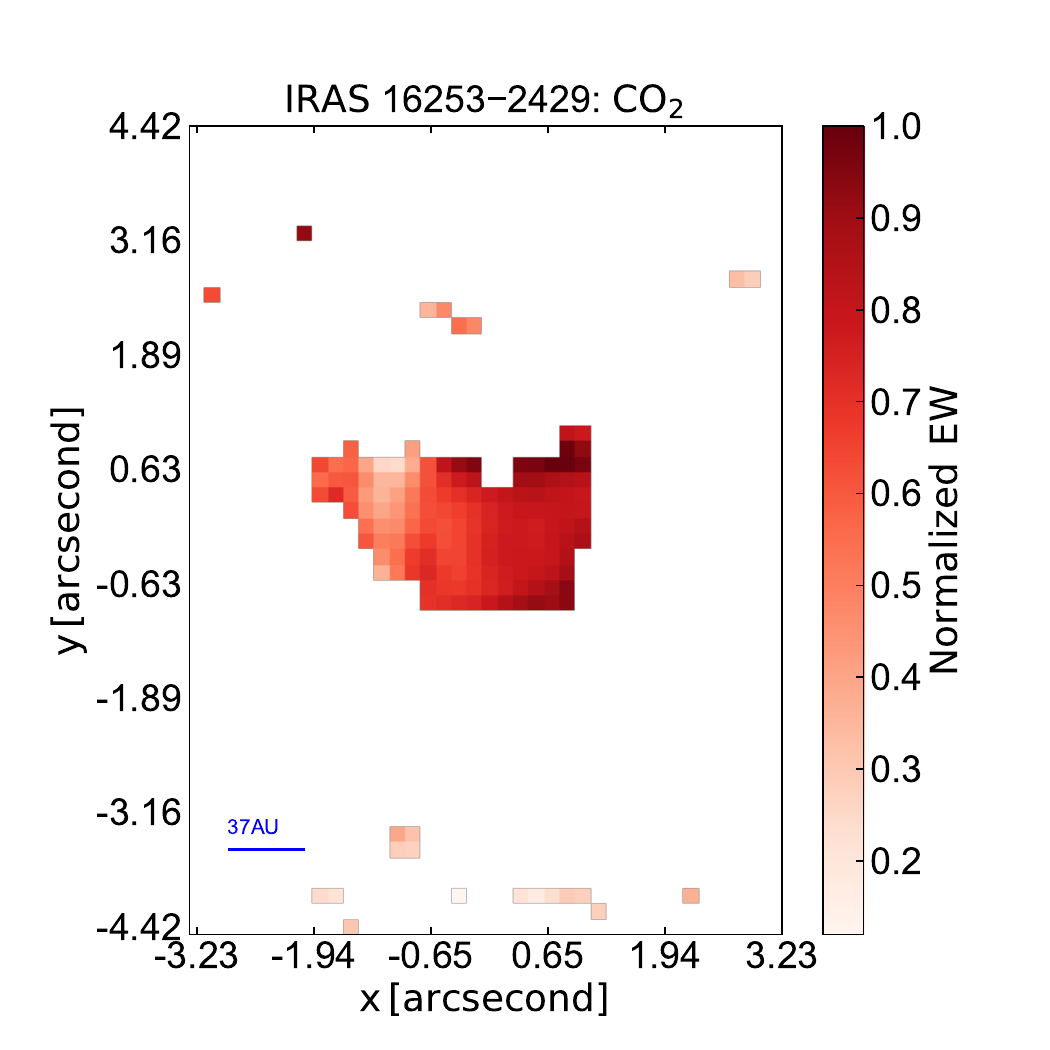}
\caption{\label{DIST} The equivalent width of the  absorption dips of $\rm{CH_4}$, $\rm{CO_2}$ from JWST MIRI/IFU observation \cite{Lei2024}. }
\label{fig:cons}
\end{figure}

Furthermore, the distribution of metabolic byproducts would trace the reactant molecules. For instance, as highlighted in Ref. \cite{Methanogens}, the distribution of methane (or acetic acid) may align with that of solid carbon dioxide. This distribution consistency has been studied in the protostellar region IRAS16253-2429, as illustrated in Fig. \ref{fig:cons}, based on observations from the James Webb Space Telescope \cite{Lei2024}. However, this distribution consistency can also be naturally explained by the "Classical" dark-cloud chemistry model, as detailed in \cite{Herbst1973,Caselli2012}. Further research is needed to determine whether a portion of the detected methane originates from methanogenic life activities.

\subsection{Microfossils in meteorites}

As discussed above, remnants or fossils of life from the pre-solar nebula may have diffused throughout the solar system. Therefore, it is reasonable to search for such remnants or fossils within objects that originated in the pre-solar nebula, particularly in primitive meteorites. However, meteorites are inevitably contaminated with terrestrial microorganisms after falling to Earth. In Ref. \cite{Tait}, the authors discovered fossilized microorganisms preserved within calcite-gypsum admixtures in meteorites. This discovery highlights the inherent uncertainties in the search for extraterrestrial microfossils within meteorite samples.

There have been numerous studies focused on searching for fossils in meteorites. The most famous discovery comes from the Martian meteorite ALH84001 \cite{McKay1996} which claims to have identified evidence of extraterrestrial life.  However, many researchers have argued that the various features observed in ALH84001 may have non-biological origins \cite{Bradley1997}. In Ref. \cite{Lin2014}, the authors utilized Nanoscale Secondary Ion Mass Spectrometry to analyze organic carbon particles found within the Tissint meteorite. Their studies indicate that the carbon isotopic compositions of the organic matter are significantly lighter than Martian atmospheric $\rm CO_2$ and carbonate, suggesting a possible biotic process or a derivation from carbonaceous chondrites. In Ref. \cite{Rozanov}, the authors present potential microfossils in the Orgueil meteorite.

It is important to clarify that identifying fossils in meteorites does not automatically confirm the existence of molecular cloud life in the pre-solar nebula, particularly in the case of Martian meteorites. Furthermore, establishing a genetic link between these extraterrestrial fossils and Earth's life forms remains an exceptionally challenging project.

\section{Summary}

Molecular cloud biology, introduced as a novel conceptual framework within this article, represents a new research branch of astrobiology. This discipline systematically investigates the potential existence, characteristics, and mechanisms of life forms within molecular clouds, with particular emphasis on elucidating their unique biophysical and biochemical mechanisms, tracing evolutionary lineages through descendant identification and fossil record analysis, and exploring other relevant scientific issues. Molecular cloud biology establishes theoretical connections with multiple scientific domains, notably low-temperature physics, origins of life research, and the search for extraterrestrial biosignatures, thereby providing a comprehensive approach to understanding life's potential in cosmic environments.

The exploration of molecular cloud biology is currently in its nascent stages, with a multitude of complex phenomena and mechanisms yet to be fully elucidated. In this draft, we aim to establish a comprehensive conceptual framework to facilitate the systematic understanding of molecular cloud biology, while also seeking to ignite intellectual curiosity and stimulate further research interest in this field.

\acknowledgments
This work is supported by the National Natural Science Foundation of China (Grants No. 12373002, 11773075) and the Youth Innovation Promotion Association of Chinese
Academy of Sciences (Grant No. 2016288).

%\clearpage

%*************************************

%\newpage

\renewcommand{\refname}{References}

\end{document}